# Investigation of ITO based liquid sensor for ammonia hydroxide detection

**Seok-hwan Lee and Jiho Chang***

*OST School, Korea Maritime and Ocean University, Busan 606-791, Korea*

**Jung-Yeul Jung**

*Division Technology of Center for Offshore Plant Industries (KRISO), Ansan 305-343, Korea*

**Moonjin Lee**

*Division of Maritime Safety Research (KRISO), Ansan 305-343, Korea*

We proposed an Indium Tin Oxide (ITO) sensor for detecting the $NH_4OH$ in seawater and carried out a series of experiments to investigate the feasibility of it as a hazardous and noxious substance (HNS) sensor. The ITO layer revealed a distinct resistance change ($\delta R$) which is linearly correlated with the $NH_4OH$ concentration. Sensing mechanism of the porous ITO layer has been explained in terms of reduction and electrical double layer (EDL) formation. Also, the chemical stability of ITO as a HNS sensor has verified.

# I. INTRODUCTION

Hazardous and noxious substances (HNS) spill accidents happen frequently around Korean peninsula. The HNS spill accidents in the ocean have been recognized as a critical problem of spoiling the ocean habitats, human health and ecosystems as well [1-2]. Unlike the oil spills, however, HNS spill issue was not getting the spotlight so far. Note that there are two basic ways of HNS detection by using spectroscopic methods and direct chemical reaction on the sensor surface. However, those ways neither provide high spatial resolution and tracking ability, although those issues are essential for minimizing the HNS spill accident damage in the ocean. One possible solution is making a floating sensor network around the dangerous area on the ocean surface. Hence, light-weight and mass-producible sensor is inevitable for developing an advanced HNS prevention system.

We proposed to use a metal oxide as a detecting material and to adopt a simple printing method. Recently, metal oxides ($SnO_2$, $TiO_2$, $WO_3$, ZnO, and $In_2O_3$) have been investigated for various sensors, since electrical conductivity of metal oxide varies immediately along with the change of environmental circumstances [3-7]. Among the various metal oxides, we have focused on the indium tin oxide (ITO). ITO is an n-type semiconductor with a wide band gap (3.5–4.06 eV) [8]. Also it has been well known as a material with an intrinsic chemical stability and high electric conductivity. Especially, chemical stability of ITO makes it a good candidate for HNS sensors, because it should operate in the harsh environments. Moreover, since one can use a powder phase ITO, simple printing process is applicable for the fabrication of advanced sensors [9].

In this study, we have investigated the feasibility of printed ITO layer as a new HNS detector. Printing procedure has investigated and the basic sensor performances were evaluated. Also robustness of the ITO layer has been tested.

# II. EXPERIMENTS

We made a paste for printing that contains the ITO powder (90% $In_2O_3$, 10% $SnO_2$, # 2731BY, manufactured by Nanostructured & Amorphous materials) and an organic binder (ethyl cellulose + $\alpha$-terpineol) with 1:1 weight ratio. We used a quartz ($SiO_2$) substrate. The size of the substrate is about $35 \times 35 \times 1$ mm$^3$. It was cleaned sequentially with methanol, acetone and DI water in an ultrasonic bath (each step took 15 minutes) to remove organics and impurities on the surface. ITO powder was printed using a zigzag-pattern mask (thickness: 30 μm, mesh size: 325 inch$^{-1}$) on a quartz substrate. After printing, the ITO layer was heated at 200 ℃ for 1.5 hour in order to remove the volatile organic solvent in the printed layer (debinding process). We have optimized the viscosity of the paste in terms of the transfer ratio which means the ratio between the linewidth of original pattern and printed one. To monitor the electric signal from the ITO layer, we made contacts at the both ends of the ITO layer using a silver paste.

Transmission electron microscopy (TEM) was used to observe the particle size distribution and crystallographic quality of the ITO particles. Field emission scanning electron microscopy (FE-SEM) was used to investigate the surface of printed ITO layer. Structural property of the ITO layer was characterized by X-ray diffraction (XRD). The sensor operation was monitored by using a current-voltage source. We have recorded the resistance of the sensor in the air as a reference signal, and monitored the variation of it by soaking in the seawater and the ammonia hydroxide solutions. We have monitored the temporal variation of resistance in the air and soaked in the solutions. We used four different liquids; DI water, seawater, 10% $NH_4OH$, and 20% $NH_4OH$ solution. Floating platform is necessary to position the sensor on the liquid surface. The ITO sensor is placed just beneath the platform. I-V characteristics were measured under a DC bias of 5V using a Keithley 2400V instrument.

## III. RESULT AND DISCUSSION

First of all, ITO powder was evaluated to verify the suitability for printing techniques. ITO particles were analyzed by transmission electron microscopy (TEM) as shown in the Figure 1(a). The ITO particles revealed well-defined shape with 20-30 nm diameters. The inset of Fig. 1(a) shows the size distribution of ITO particles. The mean diameter ($d_{mean}$) was estimated to be 38.5 nm with the standard deviation of 5.8 nm. Also, a few numbers of relatively large particles ($d_{mean} > 80$ nm) were observed. However, we thought that the commercial ITO composite can be used for printing because we used a screen with a relatively large linewidth (~1mm). Puetz and Aegerter have reported a successful screen printing result of the ITO layer [10]. They used a mask with 100 μm linewidth and 25~50 nm diameter ITO particles. In compare to their result, we considered that the mask linewidth to particle size ratio of our experiment is large enough to apply the printing process.

Using the ITO powder, ITO layer was printed on the quartz substrates. Fig. 1(b) is the FE-SEM images of the surface of printed ITO layer and the inset shows the cross-section of it. From the surface, a porous structure with large grains was observed. Porosity of the layer was estimated to be 27~30%. Note that the nominal grain size (ranging 100 – 200 nm in diameter) in the Fig. 1(b) is considerably larger than the ITO particles observed from the TEM observation (Fig. 1(a)). It indicates that the large ITO grains were formed during the debinding process. The inset of Fig. 1(b) is the cross sectional view of the ITO layer. The thickness of the layer was found to be 10 ± 1 μm and porous structure with large ITO grains was observed too. Consequently, highly porous ITO layer was obtained which is especially suitable for the sensor application.

The Fig. 1(c) is the X-ray diffraction pattern of the ITO composites and the printed ITO layer. Strong diffraction peaks were assigned to (222), (400), (440) and (622) planes of cubic bixbyite ITO (JCPDS 01-089-4198). We have mentioned that the size of ITO grain increased during the debinding process, however from the XRD result no clear evidence of crystallinity change was observed because of the low process temperature (200 $^{o}$C). It has reported that at least 350$^{o}$C is required to change the crystallinity of ITO [11].

To investigate the feasibility of the ITO layer as a HNS sensor, we have monitored the resistance change of the ITO layer in according to the various liquids as shown in the Fig. 2. First of all, to find out a reference signal level, the resistance of the ITO sensor has been monitored in the air for 50 seconds, then the ITO layer was soaked in the liquids. The resistance of the ITO layer abruptly dropped, and slowly increased to a certain value from all samples (one can observe some noise signal during the transition). Note that, the resistance change ranges in ~2 orders, also is correlated with the kind of liquid. The reference signal (initial resistance values in the air) of the ITO layer is as large as $4.0\pm0.1$ $\times10^9$ $\Omega$. Then, it dropped to its minimum value by soaking in the liquids. The minimum resistances were respectively $3.3\times10^7$ $\Omega$ (DI water), $4.9\times10^6\Omega$ (seawater), $1.1\times10^7\Omega$ (10% $NH_4OH$ solution), and $5.2\times10^7\Omega$ (20% $NH_4OH$ solution). After ~15 seconds of stabilizing time, the resistance of ITO sensors were saturated to $1.8\times10^8$ $\Omega$ (DI water), $4.1\times10^7$ $\Omega$ (seawater), $1.9\times10^7$ (10% $NH_4OH$ solution) and $1.0\times10^7$ $\Omega$ (20% $NH_4OH$ solution), respectively. Fig. 2 clearly implies that the printed ITO sensor can be used for the $NH_4OH$ detection in the seawater.

Here, we have been estimated the highest residual HNS concentration to be no more than 20%, because high HNS concentration could not be maintained for long time due to the physical and chemical dilution in the ocean[12].

The resistance change of the ITO layer (shown in the Fig. 2) can be simply explained by the redox reaction and the electrical double layer (EDL) formation. In generally, redox reactions (oxidation-reduction reactions) are simultaneous reactions, meaning they cannot occur without each other. Oxidation means the loss of electrons during the reaction and reduction is classified as the gain of electrons in a reaction. Due to the change of free electron density, the conductivity of material will be changed. Remind that when ITO layer is soaked into the solution, the resistance has dropped. It means that free electron density at the surface has increased by the reduction.

It should be noted that one should consider the effect of electrolytes in the seawater also. Thus, the net influence of ammonia itself should be expressed as the difference between the resistance of the NH$_3$ solutions and that of the seawater.

Since the reaction happens in an electrolyte, electronic double layer (EDL) will also be formed near the sensor surface [15-16]. It is well known that the EDL is consisted with two layers; the first layer (stern layer) comprises counter ions adsorbed onto the surface due to chemical interactions and the second layer (diffuse layer) is composed of ions attracted to the surface charge via the Coulomb force, electrically screening the first layer. This second layer is loosely associated with the surface. Formation of EDL will make an equilibrium state near the surface and further resistance change will be hampered as shown in the Fig. 2.

Fig. 3(a) and 3(b) reveal the responses of the ITO sensor. The sensitivity of it is defined as $S = (R_{air} - R_{liq})/R_{air} \times 100(\%)$. Where $R_{air}$ and $R_{liq}$ is the resistance of ITO sensor in the air and liquid respectively. As the NH$_4$OH concentration increasing, sensitivity increases also. Linear response was obtained up to the NH$_4$OH concentration of 15%, however, nonlinearity was observed at higher concentration region. Rout et al [14] have observed a similar phenomenon from a SnO$_2$ and In$_2$O$_3$ based NH$_3$ gas sensors. They observed nonlinear response from both SnO$_2$ and In$_2$O$_3$ gas sensors. SnO$_2$ based ones revealed comparably stronger nonlinearity and higher sensitivity as well. They argued that the nonlinear response is caused by higher carrier concentrations in the SnO$_2$. Also they mentioned that it is potentially related with the grain size. In our experiment, since we just used one kind of ITO composite, grain size effect might be negligible. Therefore, we temporally attributed the sensitivity saturation to the high intrinsic carrier concentration of ITO.

Fig. 3(b) shows the sensitivity variation along with the ambient temperature change. Note that, it is quite an important factor for the HNS sensor application, because the sea surface temperature is varying from ~ 0 to 30°C according to the position on earth, also the sensitivity of a sensor is strongly related with the ambient temperature as well.

When the ambient temperature varied from 5 to 35 ℃, sensitivity increased about 4%, which corresponds to ~5% of liquid concentration change. It implies that temperature compensation should be considered for the practical use of ITO sensor.

To evaluate the robustness of the ITO sensor, long-term operation test has performed. We have chosen a printed ZnO layer for the comparison purpose. Both samples have been soaked in the seawater at room temperature for 24 hours under the external bias (5V). Fig. 4 shows the SEM images of the (a) as-printed ITO layer, (b) tested ITO layer, (c) as-printed ZnO layer, and (d) tested ZnO layer. As one can see, significant change was not observed from the ITO layers (Fig. 4(a) and (b)), while, dramatic change of surface morphology was observed from the ZnO layers (Fig. 4(c) and (d)). The insets of each picture are corresponding XRD results of those samples. The inset of Fig. 4 (a) reveals the cubic bixbyte phase ITO with an intense (222) peak at 30.6 degree, and negligible change was obtained from the tested ITO layer (the inset of Fig. 4(b)). However, clearly opposite result was obtained from ZnO layer. The inset of Fig. 4(c) reveals a typical pattern of hexagonal phase ZnO with an intense (1000) peak at 31.7. While, significant decrease of crystallinity was observed from the tested one (the inset of Fig. 4(d)) in consistent with the result from SEM observations. This result is attributed to the chemical stability of ITO, which reveals well the suitability of ITO as a material for HNS sensor.

# IV. CONCLUSION

Feasibility of the ITO layer for the HNS detection has been investigated. The ITO composite has a nominal size distribution which is applicable for the conventional printing process. The printed ITO layer has a porous structure which is preferable for developing a sensor with high sensitivity. Sensor operation was successfully demonstrated by monitoring the temporal resistance change of the ITO layer when it is soaked into the various liquids. The ITO sensor revealed linear response up to the $NH_4OH$ concentration of 15%. Also, we have pointed out that the compensation of temperature change is required for the practical use. Robustness of the printed layer has been evaluated by a long-term operation test, which has been attributed to the chemical stability of ITO.

## ACKNOWLEDGEMENT

This study was supported by "Research & Development Program of the Hazardous & Noxious Substance system maintenance and accident management" funded by Ministry of Oceans and Fisheries of Korea. And also ~~ ( 추가 필요)

**Figure Captions.**

Fig. 1 (a) Particle size distribution of ITO composites (the insert is the TEM image of it), (b) FE-SEM images of the ITO layer (the inset is the cross section of it), and (c) XRD pattern of the ITO composites and the printed ITO layer.

Fig. 2 Time resolved resistance change of the ITO layers soaked in the various liquids.

Fig. 3 Responses of the ITO sensor. (a) $NH_4OH$ concentration dependency, and (b) temperature dependency.

Fig. 4 FE-SEM images of (a) as-printed ITO layer, (b) tested ITO layer, (c) as-printed ZnO layer, and (d) tested ZnO layer. The insets are corresponding XRD results.

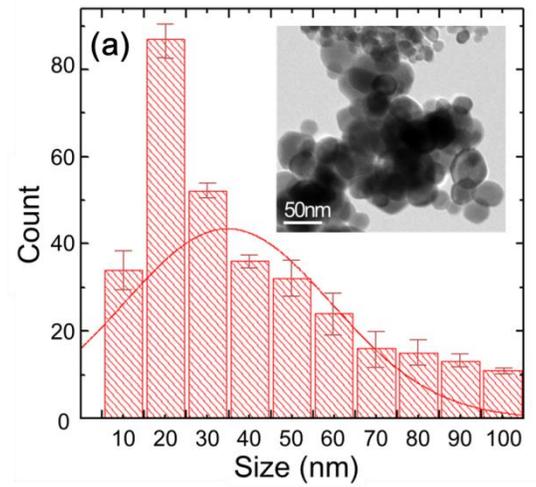

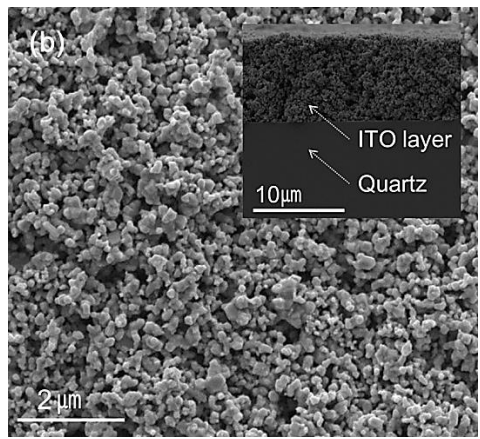

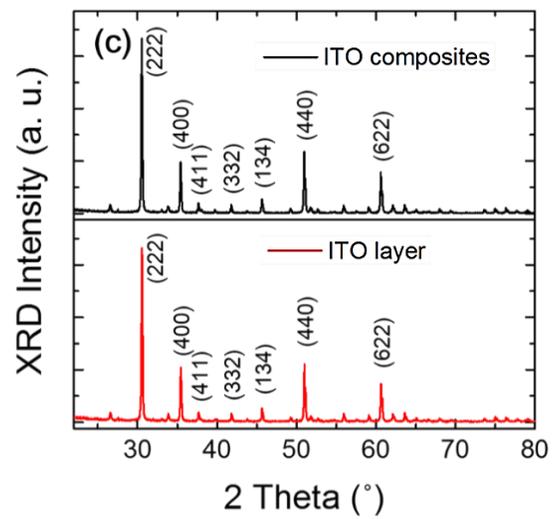

Fig.1 S.H. Lee et al.

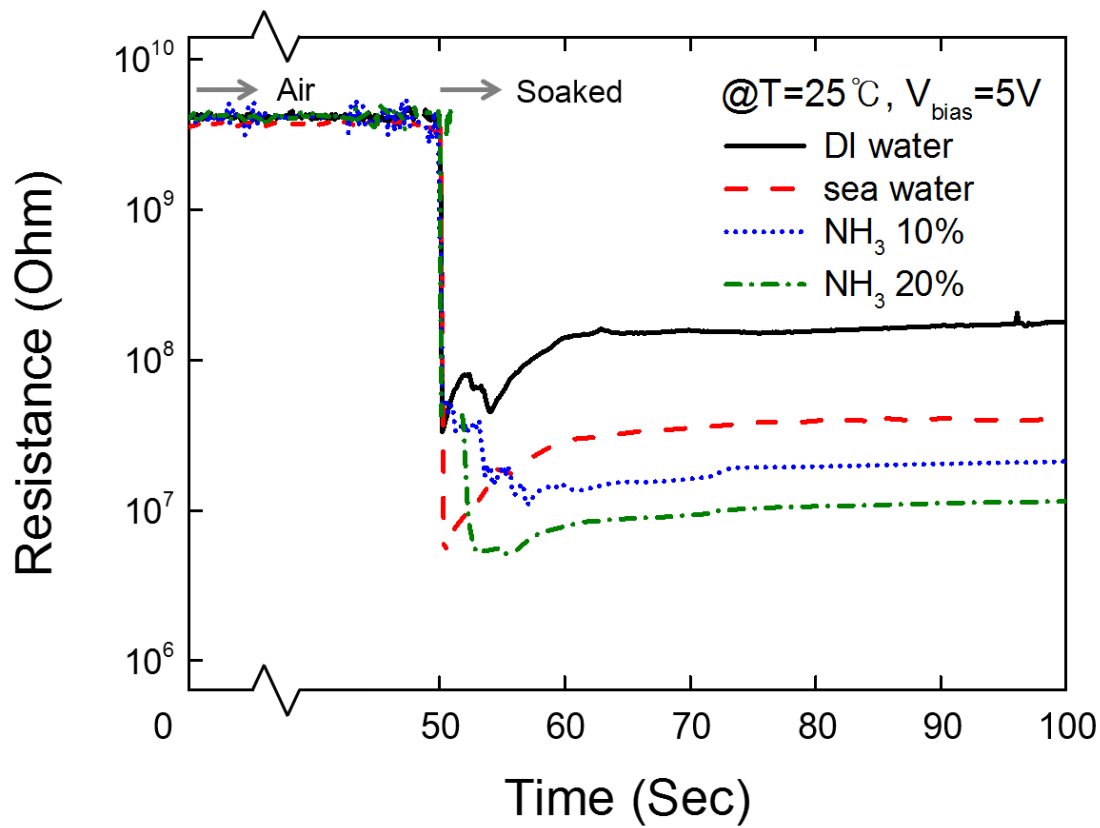

Fig. 2  S.H. Lee et al.

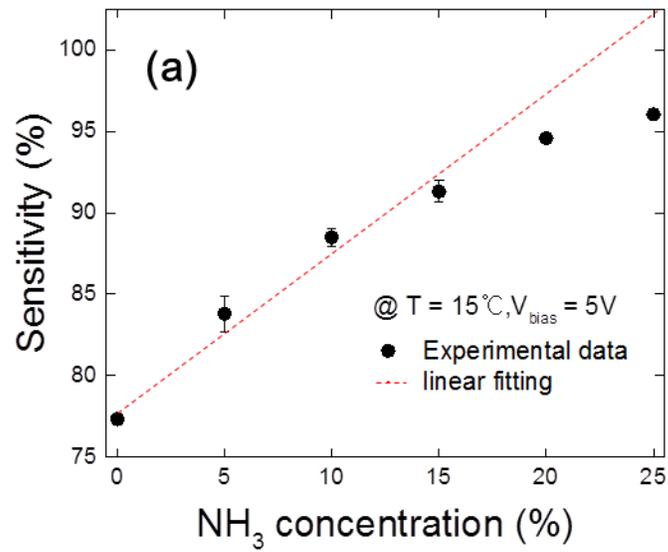

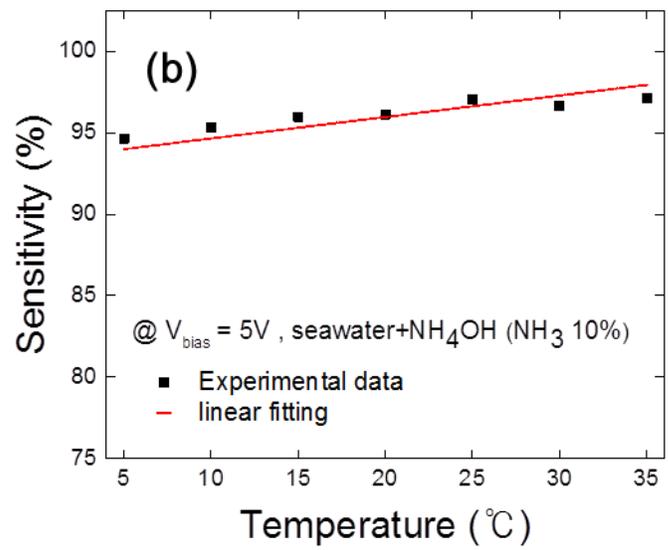

Fig. 3  S.H. Lee et al.

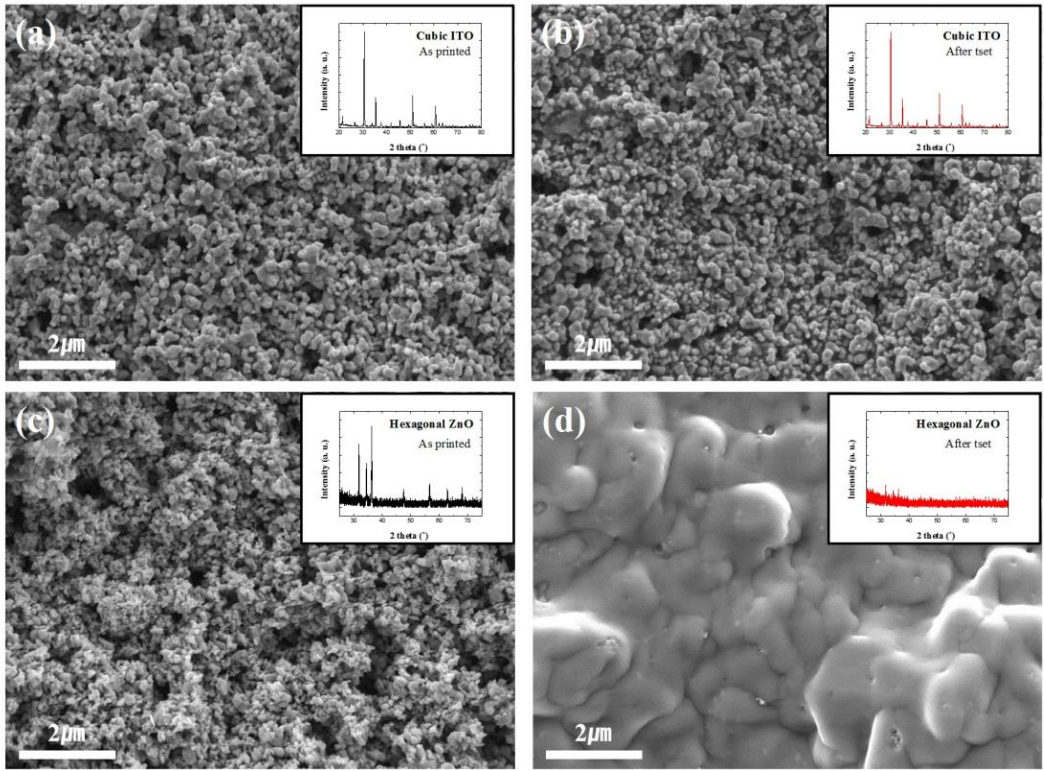

Fig. 4  S.H. Lee et al.